\documentclass[slac_one]{revtex4}
\usepackage{graphicx}
\usepackage{fancyhdr}
\begin{document}

\title{{\small{2005 ALCPG \& ILC Workshops - Snowmass,
U.S.A.}}\\ %% Please keep this conference title here
\vspace{12pt}
SUSY Parameter Determination} %% Paper title goes here


%
%\author{P. Bechtle}
%\affiliation{SLAC, Stanford, CA 94025, USA}
%\author{K. Desch}
%\affiliation{Universit\"at Freiburg, Physikalisches Institut, Hermann-Herder-Str. 3, D-79104, Germany}
\author{R. Lafaye}
\affiliation{CERN, CH-1211 Geneva 23, Switzerland}
\author{T. Plehn}
\affiliation{Max Planck Institute for Physics, Munich, Germany}
%\author{P. Wienemann}
%\affiliation{Universit\"at Freiburg, Physikalisches Institut, Hermann-Herder-Str. 3, D-79104, Germany}
\author{D. Zerwas}
\affiliation{LAL, B.P.34, 91898 Orsay Cedex, France}

\begin{abstract}

The impact of the LHC, SLHC and the ILC on the precision of the determination
of supersymmetric parameters is investigated.
In particular, in the point SPS1a the measurements performed at the ILC 
will improve by an order of magnitude the precision obtained by the LHC alone.
The SLHC with respect to the LHC has the potential to reduce the errors by
a factor two.

\end{abstract}

\maketitle

\thispagestyle{fancy}


\section{Introduction} 

The supersymmetric
extension~\cite{Wess:1974tw} of the Standard Model is a well motivated extension
providing us with a description of
physics that can be extended consistently up to the unification scale.
If supersymmetry is discovered in the next generation of collider experiments, 
it will be crucial to determine its
fundamental high-scale parameters from weak-scale
measurements~\cite{Blair:2002pg}. 
The wealth of measurements~\cite{ATLAS,Weiglein:2004hn} will require precise theoretical
predictions~\cite{Porod:2003um,Djouadi:2002ze,Allanach:2001kg} as well as complex
tools such as Fittino~\cite{Bechtle:2004pc,Bechtle:2005vt} and 
Sfitter~\cite{Lafaye:2004cn} to properly determine the underlying fundamental parameters.

In the following sections, the SPS1a~\cite{Allanach:2002nj} point is explored 
with the standard set of measurements
as detailed in~\cite{Weiglein:2004hn}, corresponding to an integrated luminosity of 
300~fb$^{-1}$ for the LHC. The impact of the SLHC, with
a luminosity ten times larger is explored. To obtain the experimental errors 
for the LHC, the statistical errors were scaled and the systematic error was kept fixed.
For the ILC a maximum of 1000~fb$^{-1}$ is used at centre of mass energies up to 1~TeV.

\section{SUSY parameter determination}

The masses measured in SPS1a of the LHC, the ILC and their combination
allow to perform a fit of the mSUGRA parameters. In particular, even if the starting point of the fit 
is far away from the true parameters (e.g. 1TeV for m$_0$ and m$_{1/2}$), the fit converges 
to the true values. The sign of $\mu$ was fixed to its true value.

\begin{table}[htb]
\caption{\label{Tab:Msugra} Results for mSUGRA for LHC (masses and edges), ILC and LHC+ILC}
\begin{tabular}{|l|r|cccc|}
\hline
            & SPS1a  & $\Delta$LHC$_{masses}$ & $\Delta$LHC$_{edges}$ & $\Delta$ILC & $\Delta$LHC+ILC\\
\hline
m$_0$       & 100    & 3.9 & 1.2 & 0.09 & 0.08 \\
m$_{1/2}$   & 250    & 1.7 & 1.0 & 0.13 & 0.11 \\
$\tan\beta$ & 10     & 1.1 & 0.9 & 0.12 & 0.12 \\
A$_0$       & -100   & 33  & 20  & 4.8  & 4.3  \\
\hline
\end{tabular}
\end{table}
The LHC (Table~\ref{Tab:Msugra} column $\Delta$LHC$_{masses}$) can determine the parameters with a precision
at the percent level, while the ILC ($\Delta$ILC) improves the determination by an order of magnitude. 
It is interesting to note that the results of the LHC can be improved (Table~\ref{Tab:Msugra} 
$\Delta$LHC$_{edges}$), significantly
by using the measured edges, thresholds and mass differences in the fit instead of the masses. 
As the masses are determined from the edges in long decay chains, 
the resulting masses are strongly correlated. In order to restore the initial
sensitivity, the full correlation matrix would be necessary.

Of 14 measurements at the LHC in SPS1a, 6 can be improved only marginally by the increase 
of integrated luminosity at the SLHC, while 8 can be improved by up to a factor~2. 
The effect on the errors on the mSUGRA parameters is shown in Table~\ref{Tab:SLHC}. 
While the errors from the SLHC alone are reduced by 50\% (with the exception of $\tan\beta$), the combined
errors SLHC+ILC are essentially the same as LHC+ILC. Only the error on A$_0$ is reduced by 20\%.
\begin{table}[htb]
\caption{\label{Tab:SLHC} Results for mSUGRA for LHC, SLHC, ILC, LHC+ILC, SLHC+ILC}
\begin{tabular}{|l|r|cccc|}
\hline
            & SPS1a  & $\Delta$LHC & $\Delta$SLHC & $\Delta$LHC+ILC & $\Delta$SLHC+ILC\\
\hline
m$_0$       & 100    & 1.2 & 0.7 & 0.08 & 0.07 \\
m$_{1/2}$   & 250    & 1.0 & 0.6 & 0.11 & 0.11 \\
$\tan\beta$ & 10     & 0.9 & 0.7 & 0.12 & 0.12 \\
A$_0$       & -100   & 20  & 10  & 4.4  & 3.8  \\
\hline
\end{tabular}
\end{table}

The precision obtained with the previous fits neglects the theoretical errors. In fact, 
if reasonable theoretical errors
such as 3~GeV~\cite{schweinlein} 
on the lightest Higgs boson, 3\% on coloured sparticles, 1\% on neutralinos and sleptons are taken into account,
the error on the m$_0$ mass increases by an order of magnitude and the other errors are 
twice as large in the combined
LHC+ILC fit. The experimental precision, especially on the ILC measurements will necessitate 
a vigorous theoretical effort~\cite{SPA} 
in order to fully exploit the available experimental information.

%When theoretical errors are taken into account, a bottom-up test can be performed: the spectrum is generated with one spectrum
%calculator and fitted with different one. 
%This exercise has been performed with SUSPECT~\cite{Djouadi:2002ze} and 
%SOFTSUSY~\cite{Allanach:2001kg}. Th results are in general satisfactory, 
%i.e., the fundamental parameters are reconstructed, 
%albeit with slightly shifted central values, in agreement with the generated parameters. Only m$_0$ 
%was significantly outside the error bar,
%but this could be due to the use of an older version of the spectrum calculator.

%\subsection{Beyond mSUGRA}

Between the two extremes of a tightly constrained model such as mSUGRA with only 5 parameters and a full fledged
MSSM with more than 120 parameters, it is interesting to study intermediate 
models~\cite{lhNonUniversal}. This approach is 
an alternative to fixing parameters in the MSSM. In particular, the 
scalar sector (m$_0$) was separated into three independent parameters: 
one for the sleptons, one for the squarks and one for the Higgs sector.
The result of the study is shown in Table~\ref{Tab:notMsugra}. 
The Higgs sector is undetermined as at the LHC only the
lightest neutral Higgs boson is observed in SPS1a which is not sensitive to m$_{H_{ud}}^2$. 
The scalar sector for the squarks is less well determined
as the quarks are heavier and their measurement less precise than those of 
the sleptons. The resulting error is proportional to the product of 
scalar mass and error.
\begin{table}[htb]
\caption{\label{Tab:notMsugra} Determination of parameters with LHC data in a non-MSUGRA model}
\begin{tabular}{|l|cccccc|}
\hline
 & m$_0^{sleptons}$ & m$_0^{squarks}$ & m$_{H_{ud}}^2$  & m$_{1/2}$  & $\tan\beta$  & A$_0$  \\
\hline
SPS1a       &  100 & 100 & 10000 & 250 & 10  & -100 \\ 
$\Delta$LHC &  4.6 &  50 & 42000 & 3.5 & 4.3 & 181   \\
\hline
\end{tabular}
\end{table}

%\subsection{Beyond SPS1a}

The determination of supersymmetric parameters is not restricted to the relatively favourable case of 
SPS1a. At the LHC even difficult regions such as the focus region can be analysed. In particular in the point
proposed by~\cite{deBoer:2004ab} the LHC will be able to measure the masses of the three neutral
Higgs bosons (in the following the heavier Higgs bosons are taken as two separate measurements 
in spite of near degeneracy). Additionally the tri-lepton
signal gives access to the mass difference between the second lightest neutralino and the lightest neutralino. 
The mass difference between gluino and $\chi_2$ will also be visible~\cite{lauren}. 
The results are shown in Table~\ref{Tab:Egret} under the assumption that the measurements will be
limited by the error on knowledge of the absolute energy scale. 
The errors are shown for experimental errors and when the theoretical and parametric  
errors are also taken into account. In this point the measurement of $\tan\beta$ 
is dominated by the heavy Higgs bosons and
the scalar mass parameter by the error on the lightest Higgs boson mass. It is interesting to note that
even though the sfermions are too heavy to be observed, the fact that m$_0$ is heavy can be established
with the caveat that the theoretical error on the Higgs mass prediction is reduced. 
\begin{table}[htb]
\caption{\label{Tab:Egret} Results for EGRET with the LHC}
\begin{tabular}{|l|c|c|c|}
\hline
          &  EGRET      & $\Delta$LHC$_{exp}$  & $\Delta$LHC$_{all}$ \\
\hline
m$_0$       & 1400    & 50  & 530  \\
m$_{1/2}$   & 180     & 2   & 12    \\
$\tan\beta$ & 51      & 181 & 350 \\
A$_0$       & 700     & 0.3 &   2   \\
\hline
\end{tabular}
\end{table}

\section{Conclusions} 

The LHC will allow a first determination of the fundamental parameters of supersymmetry of the 
order of the percent, the ILC will increase the precision by an order of magnitude.
The experimental precision will have to be matched by precise theoretical predictions in order
to fully exploit the available information. For the LHC alone, the use of edges 
instead of masses increases the precision of the determination
of the fundamental parameters $m_0$ and $m_{1/2}$. The SLHC improves on the errors by roughly a 
factor~2 with respect to the LHC, as many measurements are already dominated at the LHC 
by the systematic error.

\begin{acknowledgments}

We wish to thank the organizers for the stimulating atmosphere of 
the Snowmass conference, where part of the work outlined in this paper was 
performed and Laurent Serin for the careful reading of the document.

\end{acknowledgments}


\begin{thebibliography}{9}   % Use for  1-9  references
%\begin{thebibliography}{99} % Use for 10-99 references

\bibitem{Wess:1974tw}
J.~Wess and B.~Zumino,
%``Supergauge transformations in four-dimensions,''
Nucl.\ Phys.\ B {\bf 70} (1974) 39. 

\bibitem{Blair:2002pg}
G.~A.~Blair, W.~Porod and P.~M.~Zerwas,
%``The reconstruction of supersymmetric theories at high energy scales. ((U)),''
Eur.\ Phys.\ J.\ C {\bf 27} (2003) 263
[arXiv:hep-ph/0210058].
%%CITATION = HEP-PH 0210058;%%

\bibitem{ATLAS}
ATLAS Collaboration, Detector and Physics Performance TDR, Vol.~II, CERN/LHC/99-15;
TESLA Technical Design Report (Part 3), R.D.~Heuer, D.J.~Miller, F.~Richard and 
P.M.~Zerwas ({\it eds.}), DESY 010-11 [arXiv:hep-ph/0106315];
American LC Working Group, T.~Abe {\it et al.}, SLAC-R-570 (2001), 
[arXiv:hep-ex/0106055-58];
ACFA LC Working Group, K.~Abe {\it et al.}, KEK-REPORT-2001-11, [arXiv:hep-ex/0109166].

\bibitem{Weiglein:2004hn}
  G.~Weiglein {\it et al.}  [LHC/LC Study Group],
  %``Physics interplay of the LHC and the ILC,''
  arXiv:hep-ph/0410364.
  %%CITATION = HEP-PH 0410364;%%

\bibitem{Porod:2003um}
W.~Porod,
%``SPheno, a program for calculating supersymmetric spectra, SUSY particle  decays and SUSY particle production at e+ e- colliders,''
Comput.\ Phys.\ Commun.\  {\bf 153} (2003) 275
[arXiv:hep-ph/0301101].
%%CITATION = HEP-PH 0301101;%%

\bibitem{Djouadi:2002ze}
A.~Djouadi, J.~L.~Kneur and G.~Moultaka,
%``SuSpect: A Fortran code for the supersymmetric and Higgs particle spectrum in the MSSM,''
arXiv:hep-ph/0211331.
%%CITATION = HEP-PH 0211331;%%

\bibitem{Allanach:2001kg}
B.~C.~Allanach,
%``SOFTSUSY: A C++ program for calculating supersymmetric spectra,''
Comput.\ Phys.\ Commun.\  {\bf 143} (2002) 305
[arXiv:hep-ph/0104145].
%%CITATION = HEP-PH 0104145;%%

\bibitem{Bechtle:2004pc}
  P.~Bechtle, K.~Desch and P.~Wienemann,
  %``Fittino, a program for determining MSSM parameters from collider
  %observables using an iterative method,''
  arXiv:hep-ph/0412012.
  %%CITATION = HEP-PH 0412012;%%

\bibitem{Bechtle:2005vt}
  P.~Bechtle, K.~Desch, W.~Porod and P.~Wienemann,
  %``Determination of MSSM parameters from LHC and ILC observables in a global
  %fit,''
  arXiv:hep-ph/0511006.
  %%CITATION = HEP-PH 0511006;%%

\bibitem{Lafaye:2004cn}
  R.~Lafaye, T.~Plehn and D.~Zerwas,
  %``SFITTER: SUSY parameter analysis at LHC and LC,''
  arXiv:hep-ph/0404282.
  %%CITATION = HEP-PH 0404282;%%

\bibitem{Allanach:2002nj}
B.~C.~Allanach {\it et al.},
``The Snowmass points and slopes: Benchmarks for SUSY searches,''
in {\it Proc. of the APS/DPF/DPB Summer Study on the Future of Particle Physics
(Snowmass 2001) } ed. N.~Graf,
Eur.\ Phys.\ J.\ C {\bf 25} (2002) 113
%[eConf {\bf C010630} (2001) P125]
%[arXiv:hep-ph/0202233];
%%CITATION = HEP-PH 0202233;%%

\bibitem{schweinlein}
%%\bibitem{Degrassi:2002fi}
G.~Degrassi, S.~Heinemeyer, W.~Hollik, P.~Slavich and G.~Weiglein,
%``Towards high-precision predictions for the MSSM Higgs sector,''
Eur.\ Phys.\ J.\ C {\bf 28} (2003) 133
[arXiv:hep-ph/0212020].
%%CITATION = HEP-PH 0212020;%%

%\bibitem{Fayet:1974jb}
%P. Fayet and J. Iliopoulos,
%``Spontaneously broken supergauge symmetries and goldstone spinors,''
%Phys.\ Lett.\ B {\bf 51} (1974) 461.

\bibitem{lhNonUniversal}
S.~Kraml, R.~Lafaye, T.~Plehn, D.~Zerwas, contribution to Les Houches 2005 (in preparation).

\bibitem{SPA}
The SPA project http://spa.desy.de (in preparation).


\bibitem{deBoer:2004ab}
  W.~de Boer, M.~Herold, C.~Sander, V.~Zhukov, A.~V.~Gladyshev and D.~I.~Kazakov,
  %``Excess of EGRET galactic gamma ray data interpreted as dark matter
  %annihilation,''
  arXiv:astro-ph/0408272.
  %%CITATION = ASTRO-PH 0408272;%%

\bibitem{lauren}
P.~Gris et al., contribution to Les Houches 2005 (in preparation).


\end{thebibliography}
\end{document}